# Systems Perturbation Analysis of a Large Scale Signal Transduction Model Reveals Potentially Influential Candidates for Cancer Therapeutics

Bhanwar Lal Puniya[1], Laura Allen[2], Colleen Hochfelder[3], Mahbubul Majumder[2], Tomáš Helikar[1*]

[1]Department of Biochemistry, University of Nebraska-Lincoln, NE, USA
[2]Department of Mathematics, University of Nebraska at Omaha, NE, USA
[3]Albert Einstein College of Medicine, NY, USA

**\* Correspondence:** Dr. Tomáš Helikar, Department of Biochemistry, University of Nebraska-Lincoln, Beadle Center, 1901 Vine st, Lincoln, Nebraska, 68508, USA.

thelikar2@unl.edu

## Abstract

Dysregulation in signal transduction pathways can lead to a variety of complex disorders, including cancer. Computational approaches such as network analysis are important tools to understand system dynamics as well as to identify critical components that could be further explored as therapeutic targets. Here, we performed perturbation analysis of a large-scale signal transduction model in extracellular environments that stimulate cell death, growth, motility, and quiescence. Each of the model's components was perturbed under both loss-of-function and gain-of-function mutations. Using 1,300 simulations under both types of perturbations across various extracellular conditions, we identified the most and least influential components based on the magnitude of their influence on the rest of the system. Based on the premise that the most influential components might serve as better drug targets, we characterized them for biological functions, housekeeping genes, essential genes, and druggable proteins. The most influential components under all environmental conditions were enriched with several biological processes. The inositol pathway was found as most influential in the case of inactivating perturbations, whereas the kinase pathway and small lung cancer pathway were identified as the most influential under activating perturbations. Most influential components under activating perturbations were enriched with essential genes when compared to the least influential components. Also, the most influential components were enriched with druggable proteins. Moreover, known cancer drug targets were also classified in influential components based on the affected components in the network. Additionally, the systemic perturbation analysis of the model revealed a network motif of most influential components which affect each other. Furthermore, our analysis predicted novel combinations of cancer drug targets with various effects on other most influential components. We found that the combinatorial perturbation consisting of PI3K inactivation and overactivation of IP3R1 can lead to increased activity levels of apoptosis-related components and tumor suppressor genes, suggesting that this combinatorial perturbation may lead to a better target for decreasing cell proliferation and inducing apoptosis. Lastly, our results suggest that systematic perturbation analyses of large-scale computational models may serve as an approach to prioritize and assess signal transduction components in order to identify novel drug targets in complex disorders.

## 1. Introduction

Recent advances in systems biology and computational biology have introduced methods for the visualization, comprehension, and interpretation of big data in biomedical research. These fields provide an array of methodologies including computer simulations that can be used to generate new hypotheses and identify which hypotheses might be more productive to undertake experimentally, and eliminate hypotheses with little chance of success (Kitano, 2002a;b;Ghosh et al., 2011). These methods can be effective in navigating complex network problems associated with diseases. Many diseases and pathologies can be characterized by the dysregulation or dysfunction of multiple molecular components that are connected within these highly intertwined biological and biochemical networks (Loscalzo and Barabasi, 2011). Biological networks, including biochemical signal transduction networks, consist of a large number of highly interconnected pathways that give rise to complex, non-linear dynamics governing various cellular functions (Helikar et al., 2008;Helikar and Rogers, 2009). Disruptions of these networks, such as mutations or disease states can have drastic effects upon the whole system. These effects are difficult to predict from static network diagrams.

However, understanding the hierarchy of these changes remains a paramount problem. Often the specific causal interactions of the disease state are hidden within the massive cell-wide alterations, making attempts to reverse a disease state more challenging. In addition, the specific causal interactions are difficult to predict making the development of a potential therapeutic target results in unforeseen side effects (Singh and Singh, 2012). The unwanted effects of these drugs are often drastic as seen with many cancer medications (Kayl and Meyers, 2006;Lotfi-Jam et al., 2008;Singh and Singh, 2012). Therefore, it is necessary to gain a systems level understanding of the components associated with the disease states.

In recent years, targeted therapy has been used for multiple diseases, e.g. cancer (Vanneman and Dranoff, 2012), and often involve the activation or inactivation of a specific component in a biological network by a small molecule or drug, for instance. Perturbation analyses allow one to interrogate the structure and dynamic footprint of the underlying biological system. Perturbation biology has been proposed as an approach to reduce the collateral damage caused by nonspecific drugs. Computational network perturbations and new methods to evaluate the robustness of a given network can help identify more effective network components to target in order to obtain desired outcomes with minimal disruption to the rest of the network (Molinelli et al., 2013).

In order to fully leverage the potential of computational network perturbation analyses large dynamical models are necessary. A wide spectrum of modeling approaches exists ranging from detailed (but less scalable) differential equation-based systems to large (but not dynamic) static networks. In the middle are approaches such as logical modeling that are relatively scalable while capable of capturing the dynamic nature of biological systems (Le Novère, 2015). Logical networks, namely Boolean networks, have been used to describe and simulate a wide spectrum of biological systems ranging in size as well as contextual application (Naldi et al., 2010;Helikar et al., 2012;Madrahimov et al., 2013;Rocha et al., 2013;Conroy et al., 2014).

Herein, we present results from a system-wide perturbation analysis of a large-scale Boolean model of a signal transduction network widely present in many types of cells. Specifically, the model (previously described in (Helikar et al., 2008)) represents signaling events within the integrated epidermal growth factor, G-protein coupled receptor, and integrin signaling network. The model consists of 139 components (mostly proteins) and 557 biochemical interactions. The simulation-based, system-wide perturbation analyses enabled us to identify the most and least influential components (ones with the most and least impact on the rest of the network). To explore the role and effects of these perturbations in the context of the complex extracellular environment, the simulations and analyses were conducted under four biologically relevant environmental conditions known to stimulate cell growth, cell death, motility, and quiescence (in addition to a set of randomly generated environmental stimuli). In order to investigate potential therapeutic targets, we performed functional annotation and analysis of the most influential signal transduction components under both inactivating (e.g. knock-out) and activating (e.g. over-expression) perturbations. The most influential components were found to be enriched with many biological processes and druggable targets. Also, the most influential components under activating perturbations were enriched with more essential genes than the least influential components. We identified a network of the most influential components consisting of drug targets considered in multiple cancer types. The highest ranked among the most influential components were already explored as drug targets against cancer. We used all the most influential components and their upstream regulators to identify novel interactions. We used this approach to identify novel drug targets in the signal transduction network. As a result of the systemic analysis, we identified one novel combinatorial target, PI3K-IP3R1, with consistent occurrence in all simulated environmental conditions. We simulated the effect of combinatorial perturbation and the results were correlated with the literature, further supporting our predictions.

## 2. Materials and Methods

### 2.1. Computational Model

The computational model analyzed in this work is a Boolean model of signal transduction in a generic cell type previously described in (Helikar et al., 2008). The signal transduction model was constructed manually from around 500 published papers. The model consists of several main signaling pathways, including the receptor tyrosine kinase (epidermal growth factor receptor), G-protein coupled receptors (G-alpha *i*, G-alpha *q*, G-alpha *s, and* G-alpha 12/13*)* and the integrin signaling pathways. Each component in the model corresponds to a signaling molecule (mainly protein). The model also contains nine external components that represent the extracellular environment (mostly composed of receptor ligands). It is fully annotated and freely available for simulations and/or download via the Cell Collective software (Helikar et al., 2012;Helikar et al., 2013) at www.thecellcollective.org.

Each model component can assume an active (1) or inactive (0) state at any time *t*. The activity state of the model's internal components is determined by the regulatory mechanisms of other directly interacting components. These regulatory mechanisms are described with Boolean functions (in the form of truth tables or Boolean expressions). To represent the milieu of stimuli in the extracellular environment, the model contains external components that represent various ligands. The activity level of these components is specified as a probability to simulate different levels of concentrations. This methodology was previously detailed and exemplified in (Helikar et al., 2008;Helikar and Rogers, 2009;Helikar et al., 2012;Todd and Helikar, 2012).

### 2.2. Model Simulations

The Cell Collective platform was used to perform all computational simulations of the model. Although the model is built by using discrete mathematics the output activity levels (AL) can be continuous (ranging from 0 to 100) as previously described in (Helikar et al., 2008;Helikar and Rogers, 2009). Each simulation is synchronous and consists of 800 steps, where the activity level of the measured output component is calculated as the fraction of ones (active states) over the last 300 iterations that describe the network's steady behavior (Helikar et al., 2008;Helikar and Rogers, 2009).

$$AL_{ij} = \frac{1}{NT} \sum_{j=1}^{N} \sum_{i=1}^{T} x_j(t)$$

The model was simulated and analyzed under four biologically-relevant environmental conditions: cell growth, cell death, quiescence, motility (and randomly generated conditions). Each environmental condition was defined by different combinations of the activity levels of external components (ligands). The activity level ranges of the environmental conditions were determined by an optimization method whereby 2,000 simulations were run with all external stimuli ranging from 0-100 (except for IL1_TNF and Stress that were limited to low activity levels). Subsequently, specific environmental conditions for cell growth, cell death, motility, and quiescence were determined by selecting for environmental conditions that met the activity levels of biologically relevant model components: Akt, Erk, Cdc42, Rac (Helikar et al., 2008). From the initial 2,000 random simulations, environmental conditions that yielded the appropriate biological response were averaged and the inclusion of all environmental conditions within one standard deviation created an activity range for each of the environmental components (Table 1).

A wild type (WT) experiment was conducted under each environmental condition without any perturbations. Subsequently, systematic perturbation experiments were conducted under each condition, whereby each component of the model was constitutively activated (gain-of-function/over-expression) or inactivated (loss-of-function/knock-out). Each experiment consisted of randomly selecting 100 combinations of activity levels of the external stimuli from each condition activity range. (The only exception was the random environmental condition, which was simulated 2,000 times.) Each of the 100 combinations were simulated 30 times (i.e., 30 replicates) to ensure consistency of the dynamics in response to a specific combination of stimuli. These replicates were subjected to a Fligner Killeen test of homogeneity of variables which confirmed that the measured activity levels of the network components were homologous for identical combinations of activity levels of the environmental stimuli.

### 2.3. Model Analysis

The Kolmogorov-Smirnov (KS) test (Wang et al., 2003) was used to compare the WT dynamics (under each environmental condition) with the dynamics of each perturbation experiment. If the KS test resulted in a p-value less than 0.05, then it has a difference value (DV) equal to the test statistic; otherwise, the DV for a component is zero. In order to avoid skewed results from the perturbation itself, its difference value from the WT is set to zero.

### 2.4. Most and Least Influential Components

The most influential components are defined as components that induce the largest changes in the network under a given perturbations. The ranking of the perturbations is derived by calculating an influence score (IS), which is found by the summing of each DV for all the components per perturbation experiment. The top ten percent are considered most influential, and the bottom components with IS value 0 were considered the least influential.

$$IS_i = \sum_{j=1}^{130} DV_{ij}$$

$$i = 1..130$$

### 2.5. Most Affected Components to a Specific Perturbation

For each perturbation induced, the components that are most sensitive to that perturbation are ranked in decreasing order to be able to characterize downstream effects of the perturbation on the network.

**2.6. Annotation and Biological Relevance of Signal Transduction Components**

All model components were first annotated using the appropriate NCBI gene IDs (Pruitt et al., 2007) for associated genes and UniProt IDs (Consortium, 2011) for protein products of the genes. All components were then further characterized using online resources such as DrugBank (Wishart et al., 2006).

The biological process enrichment analysis of the most influential components was done using DAVID (Huang et al., 2008), with high stringency. Gene Ontology (Ashburner et al., 2000), SP_PIR keywords, and KEGG pathways (Kanehisa, 2002) were obtained using FDR > 5%.

Essentiality data were obtained from the Online GEne Essentiality (OGEE) database and mapped on the most and least influential components (Chen et al., 2012). DrugBank data were used to obtain druggability information for each component in the network. Data on cancer associated genes were obtained from The Cancer Genome Atlas (TCGA) (Weinstein et al., 2013) and mapped on the most influential components to identify cancer associated most influential components. The enrichment of essential genes and druggable proteins was computed based on the number of genes mapped on most or least influential components out of the total number of most and least influential components.

**3. Results**

**3.1. System-wide Perturbation Analysis Reveals Core Components of the Signal Transduction Network**

A critical objective of biomedical research is the identification and prioritization of novel therapeutic targets. In this context, we performed systematic perturbation analysis in a generic signal transduction model. The workflow used in this work is illustrated in Figure 1.

The activating/inactivating perturbation experiments for each component in the model were carried out across four environmental conditions (as described in Methods). Additional randomly generated extracellular conditions were used to check the robustness of the model and results. Perturbation analysis enabled us to identify and rank components of the signaling network that are most and least influential (Supplementary Table 1). The heatmaps for all the environmental conditions (Supplementary Figures S1-S10) indicate that a few components had high influence on rest of the system. Therefore, we considered the top 10% of the components from each condition as the most influential. In contrast, the components that had no influence on the system were considered as the least influential (KS=0).

Also, the most influential components correspond to network components that, when perturbed, affect the largest part of the network in terms of the number of affected components and the magnitude of the effect. The most influential components were found for both inactivating (Figure 2 A) and activating (Figure 2 B) perturbations under the different environmental conditions. It is interesting to note that many of the most influential components overlap across all environmental conditions. However, the most influential components do not overlap between two types of perturbations (inactivating or activating).

### 3.1.1. Housekeeping Genes are Enriched in the Most Influential Components Common in Different Environments

Next, we investigated if the overlapped most influential components among different environmental conditions, have constitutive expression. Under inactivating perturbations, out of the seven components common among the different environmental conditions, PI4K, PI5K, ARF and PI3K were associated with housekeeping genes (Eisenberg and Levanon, 2013). Under activating perturbations, Trafs, Erk, Mek and SHP2 (out of nine common components), were associated with housekeeping genes. Housekeeping genes associated with the common components are displayed in Table 2. This observation suggests that the most influential components that are common among different environmental conditions are likely to function as housekeeping genes.

### 3.1.2. Unique components associated with each environmental condition are found to be condition specific

Under both types of perturbations, certain environmental conditions had several uniquely associated components (Figure 2, Table 3). Under inactivating perturbations, components uniquely associated with the cell death condition are Calmodulin (CaM), RGS, and Palpha_iR. Out of these, CaM and RGS have been previously associated with cell death and apoptosis (Fisher, 2009;Berchtold and Villalobo, 2014). In fact, CaM plays a central role in the regulation of several cellular functions, including programmed cell death (Berchtold and Villalobo, 2014). It is also known that RGS protein can regulate cell death, cell cycle and cell division (Fisher, 2009). Under activating perturbations, the most influential components associated with the cell death-inducing condition include Gbg_i and Alpha_iR. On the other hand, PP2A was found to be most influential under the growth condition, Ras and Sos under motility condition, and PAK under quiescence condition. These results are also further supported by published studies that reported Gbg_i (GNB) to be involved in mTOR-mediated anti-apoptotic pathways; Gbg_i was also functionally annotated with apoptosis and cell death (Wazir et al., 2013). PP2A was reported as a highly regulated Ser/Thr phosphatase involved in cell growth and signaling (Janssens and Goris, 2001). In pancreatic cancer cell lines, the knock down of KRAS has been found to lead to the decrease in cell motility and proliferation (Rachagani et al., 2011;Birkeland et al., 2012). Furthermore, the Grb2-Sos1 complex has been found to most likely promote cell motility, and tumerogenesis (Qu et al., 2014). These observations suggest that the proteins which were uniquely associated with simulated environmental conditions are most likely to have the association with that condition. Finally, the literature evidence obtained for housekeeping, or condition associated genes, further supports our simulation results.

### 3.2. Key Biological Processes are Enriched in the Most Influential Components

Next, we assessed the enrichment of biological processes or pathways in the most influential components. The most influential components across all four conditions under both types of perturbation showed significant enrichment with key biological processes. The counts and fold differences of enriched biological terms in all the conditions are shown in Figure 3 and 4. In the case of inactivating perturbations, inositol phosphate metabolism was enriched under all environmental conditions (Figure 3). In the case of activating perturbations, the significantly enriched biological processes include phosphate metabolic processes, kinase activity, apoptosis and interestingly, the non-small lung cancer pathway (Figure 4). These results illustrate that the group of proteins with similar biological functions appear as the influential components under each type of perturbation.

## 3.3. The most Influential Components under Activating Perturbations are Enriched with Essential Genes

Mutations in an essential gene can be lethal. Based on the hypothesis that the influential components might serve as essential for the survival of the cell, we performed essentiality analysis. Under activating perturbations, more essential genes were found within the most influential components than the least influential components (Figure 5 A). Under the cell death environmental condition, a total of 69% of the most influential components were essential; this is in contrast to the least influential components that contained 31% essential genes. Under other environmental conditions- growth, motility and quiescence, the difference of essential genes between the most influential and the least influential components are 23%, 15%, and 32% respectively.

On the other hand, under inactivating perturbations we found either an equal or larger number of essential genes in the least influential components (Figure 5 B). The most significant differences were observed under the cell death condition: the least influential components have 66% of essential genes in contrast to the 46% essential genes in the most influential. Also, under the growth condition 68% and 53% of essential genes were contained within the least and the most influential components, respectively. Under the motility and quiescence conditions, there were 3% and 9% more essential genes within the least influential components than the most influential components, respectively. We found that under inactivating perturbations, the number of essential genes among the least influential components were slightly larger than the activating perturbation (Figure 5 C and Figure 5 D). On the other hand, under activating perturbations, the more essential genes mapped within the most influential components than the least influential components.

Thus, the most influential components are essential under activating perturbations, suggesting an environmental condition-specific essentiality.

## 3.4. The Most Influential Components are enriched with Druggable Proteins

To further investigate the importance of the most influential components, we evaluated the distribution of known druggable targets. We obtained druggability data from the DrugBank database (Wishart et al., 2006) and mapped them on the most and least influential components. A total of 51 components in the whole network were enriched with druggable proteins. We found that under both types of perturbations and across all environmental conditions more druggable proteins were found among the most influential than the least influential components (Figure 6). The enrichment for druggable proteins within the most influential components implicates these as critical network components.

## 3.5. The Most Influential Components as Drug Targets

### 3.5.1. Ranked most influential components based on downstream components

We identified the most affected components of the most influential components under both types of perturbations. We combined all environmental conditions to construct networks of the most influential components with their downstream targets. We subsequently mapped druggable proteins and cancer associated genes on these networks. Under inactivating perturbations, we obtained a network consisting of the most influential components: PI3K, EGFR, PP2A, GRK and CaM (Figure 7 A). Under activating perturbations, we obtained a network composed of influential components: EGFR, IL1_TNFR, ERK, SHP2, RKIP, Ras, Gbg_i, Fak, Integrins, and PP2A (Figure 7 B).

Total number of downstream targets for each of the most influential druggable component under both inactivating and activating perturbations are listed in Table 4. We observed that EGFR, a validated cancer drug target (Mendelsohn, 2001), affects the largest number of components under activating and inactivating perturbations.

### 3.5.2. The Most influential components mainly affect other most influential components

Here, we identified all components that directly affect the activity of each most influential component (KS=1). Interestingly, most of these direct upstream components were also ranked as the most influential in at least one environmental condition (Figure 8). Under inactivating perturbations, 22 components were directly upstream of the most influential components. Of these, 19 were the most influential under at least one environmental condition. On the other hand, under activating perturbations, out of 45 upstream components, 19 were also ranked as most influential. Additionally, under inactivating perturbations, nine (CaM, EGFR, Gbg_i, GRK, IP3R1, PP2A, PI3K, Ras, and Src) out of total 22 upstream components are druggable. Out of these 22 components, six components (CaM, EGFR, Gbg_i, GRK, IP3R1, and PP2A) were upstream to the most influential druggable components. Under activating perturbations, 21 (CaM, Cdc42, EGFR, Erk, Fak, Gbg_i, Grb2, GRK, IL1_TNFR, Integrins, IP3R1, PDK1, PI3K, PKA, PP2A, Rac, Raf, Ras, RKIP, SHP2, and Src ) out of 45 upstream to the most influential components are associated with druggable proteins. Out of these 21, ten components were also the most influential. Under both types of perturbations, a total of 18 (alpha_iR, ARF, B_Arrestin, Ca, CaM, EGFR, Gbg_i, GRK, IP3R1, Palpha_iR, PI5K, PIP2_45, PIP3_345, PP2A, RGS, PI3K, Ras, Src) upstream components were common. Nine of these components (CaM, EGFR, Gbg_i, GRK, IP3R1, PP2A, PI3K, Ras, and Src) were druggable or these were used as the drug targets. The important drug targets such as EGFR, PI3K, Ras, Raf also appeared as influential upstream components. Together, these results suggest that under inactivating perturbations the activity of the most influential components are likely to be modulated by the other most influential components.

### 3.5.3. The Most influential Components as Drug Targets, and Drug Resistance

The top most influential components such as EGFR, PI3K, ERK, and Ras etc. have been previously explored as drug targets in multiple cancer types. However, it is also evident from literature that several most influential components have been associated with drug resistance. For example, in non-small cell lung cancer, mutation within the kinase domain of EGFR, and epithelial–mesenchymal transition are responsible for the development of resistance to gefitinib (Holohan et al., 2013). In colorectal, and head and neck cancers, KRAS mutation, EGFR-S492R mutation, and increased ErBb signaling are responsible for resistance against Cetuximab (Dienstmann et al., 2012;Holohan et al., 2013). Furthermore, PI3K showed drug resistance in breast cancer against rapamycin through the expression of RSK3 and RSK4 (Rodon et al., 2013). Mutations in ERK1 or ERK2 have shown resistance against ERK inhibitors or RAF/MEK inhibitors (Wagle et al., 2014). Tumors with mutation in BRAF V600E can adapt to the RAF inhibitors (Lito et al., 2013;Perna et al., 2015). As such, the identification and prediction of drug targets alone is not sufficient to identify completely useful drug targets.

### 3.6. Regulatory Interactions between the Most Influential components and their Upstream Components

To develop a better strategy that can account for drug resistance of the most important drug targets, we sought to investigate novel regulatory interactions. We analyzed the previously described interactions between the most influential components and their direct upstream components. We found that some interactions consistently occur in more than one environmental condition. For example, the inactivation of IP3R1 increases the activity of PI3K under all four environmental conditions. However, the maximal effect was observed under the death environmental condition. Additionally, the inactivation of IP3R1 leads to inactive RGS under three conditions: cell growth, motility, and quiescence. These finding also correlate with published studies that found that RGS positively regulates apoptosis (Fisher, 2009). Other examples of consistently occurred interactions include: the activation of Grb2 leads to increased activity levels of Ras under all four environmental conditions, and increased Sos activity under death and quiescence conditions. The activation of Rac increases the activation of PAK under cell death and growth conditions. Overall, we found three types of interactions: inactivation of one component leads to the increase of activity of another component (PI3K-IP3R1, IP3R1-PI3K, RGS-IP3R1), inactivation of a component leads to decreased activity of another component (IP3R1-RGS), and activation of a component leads to increased activity of another component (Grb2-Ras, Grb2-Sos, Rac-PAK).

The fold differences of all these interactions are displayed in the Table 5. Under the cell death condition, the inactivation of IP3R1 results in PI3K activity increase by 2.38 fold. Similarly, PI3K inactivation leads to a 5.42 fold increase in IP3R1 activity. In the case of other interactions, the inactivation of IP3R1 leads to inactive RGS under the cell growth, motility and quiescence conditions. Under the motility and quiescence conditions, the inactivation of Gbg_i leads to inactive CaM. The activation of Grb2 increases the activity of Ras 7.40 fold under the cell death condition, and 2.13 fold under the quiescence condition. Grb2 activation also affects Sos 7.8 fold under the cell death condition and 2.18 fold under the quiescence condition. An activating perturbation of Rac increases the activity of PAK more than 18 fold under the cell death condition, and 5.59 fold under the growth condition.

These results suggest different types of regulatory effects of activating and inactivating perturbations of direct upstream components of most influential components.

**3.7. Co-targeting IP3R1 with PI3K**

As discussed earlier, although PI3K was identified as one of the most influential components, it has been also associated with drug resistance. Based on the interactions of upstream regulators of most influential components discussed above, we further investigated the interactions involving PI3K and IP3R1 with the objective of identifying a secondary drug target that could be potentially used to address the issue of PI3K-associated drug resistance. In contrast to PI3K/Akt signaling, IP3R1 positively regulates apoptosis. We hypothesized that the rate of apoptosis will increase when IP3R1 is overactivated (activating perturbation) and PI3K is inactivated (inactivation perturbation). Despite the strong dynamical relationship between IP3R1 and PI3K, these two components are only connected indirectly through a sub-network. In this sub-network, Gbg_i is upstream of and directly activates both components. IP3R1 regulates PI3K through a Ca->EGFR route, whereas PI3K regulates IP3R1 via a PTEN -> PIP2_45 -> IP3 route (Figure 9).

The inactivating perturbation of PI3K resulted in the inactivation of 29 components across all four environmental conditions. Under PI3K inactivation, the average activity of IP3R1 increased from 71.9% in wild type to 85.18%. This perturbation also led to down-regulation of positive regulators of apoptosis phospholipase A2 (PLA2) and arachidonic acid (AA). AA released by PLA2 triggers Ca2+ dependent apoptosis through mitochondrial pathways (Penzo et al., 2004). The elevation in Ca2+ is thought to be involved in apoptosis (Pinton et al., 2008). It was shown that blocking calcium channels can directly lead to tumor promotion (Mason, 1999). Thus, inactivation of PI3K can block cell proliferation; simultaneously it can lower the rate of apoptosis.

Under growth condition, the activating perturbation of IP3R1 increased the activity of apoptosis-associated components: Ca, CaM, CaMK, CaMKK and RGS in the range of +1.41 Fold to +2.09 fold when compared to wild type.

To simulate the cell death effect under growth condition, we carried out a double perturbation of IP3R1 and PI3K whereby we constitutively activated IP3R1 and inactivated PI3K1. Under this combinatorial perturbation, we found 27 proteins including proto-oncogenes such as Akt (which suppresses apoptosis) and Raf to be down-regulated. Here, we found eight proteins with more than 19% increased activity than in the case of a single inactivating perturbation of PI3K. These proteins include Rap1 (+ 1.19 Fold), Ca (+ 1.21 Fold), CaM (+1.21 Fold), CaMKK (+ 1.21 Fold), Myosin (+ 1.22 Fold), CaMK (+ 1.65 Fold), PLA2 (+ 1.98 Fold), and AA (+1.98 Fold) (Table 6; full list of all affected components is given in Supplementary Table 2). It is noteworthy that these components have been found to be associated with apoptosis or cell death. Therefore, under combinatorial perturbations, components involved in cell proliferation were downregulated through the inactivation of PI3K, and the activity of tumor-suppressor genes (PLA2) with arachidonic acid (AA) and other components, including Ca, CaM, and CaMK, was increased as a result of the IP3R1 overactivation.

Together, these results suggest a novel regulatory interaction between PI3K and IP3R1, and that co-targeting both of these components may serve as therapeutic strategy rather than targeting PI3K alone.

## 4. Discussion

We have presented a systemic perturbation analysis of a signal transduction network model to identify and characterize functionally important components. We used these components to explore novel therapeutic strategies against cancer. Specifically, we used a logical modeling approach to analyze the dynamics of a large-scale signal transduction model. Logical modeling approaches have been used, for example, to understand the dynamics of signal transduction and gene regulation networks to identify drug synergies in gastric cancers, and to identify potential drug combinations (Flobak et al., 2015). In biochemical networks, combined effect of topology and dynamical features have been shown to have the most significant impact on the dynamics of the network (Kochi et al., 2014). Computational approaches have become indispensable tools to understand biological pathways and disease phenotypes. Examples include computational methods such as molecular modeling, text mining, and network modeling to identify drug targets in a vast array of diseases from pathogens to complex disorders (Flórez et al., 2010;Yao et al., 2010;Folger et al., 2011;Madrahimov et al., 2013;Puniya et al., 2013).

In the present work, the identified most influential components were characterized for biological functions. The relevance of identified influential components was established with pathway analysis, mapping of housekeeping genes, essential proteins, and association with druggable proteins. Interestingly, we found enrichment of housekeeping genes in the most influential components that were independent of the extracellular environments. A notable agreement is obtained from literature surveys for the most influential components, which were unique to specific environmental conditions. Because essential components are important from a disease perspective, the identified most influential components may serve as potential candidates and essential proteins under specific conditions. Under activating perturbations, we found that essential genes were enriched more within the most influential components than within the least influential components. The high association of dysregulated signal transduction proteins with different subtypes of cancers suggests that these components may be important candidates for drug targets. Notably, the most influential components are enriched with several already known drug targets. However, many of these drug targets (EGFR, ERK, Ras, PI3K etc.) have been associated with drug resistance (West et al., 2002;Kobayashi et al., 2005;Linardou et al., 2008;Wheeler et al., 2010;Dienstmann et al., 2012). The mechanism of drug resistance includes mutation in the targeted protein or expression of other genes (altered expression) to bypass the effect caused by perturbation, and deregulation in apoptosis, etc. (Gottesman, 2002;Holohan et al., 2013).

Thus, to identify novel regulatory interactions, we explored components that are upstream to the most influential components associated with drug resistance. Interestingly, several upstream components (more than 90% in the case of inactivating perturbations) to the most influential components were also identified as most influential. Thus, the most influential components form a tightly connected sub-network of proteins interacting with each other. In yeast, it has previously shown that the essential proteins are hubs in the network and have more interconnections than non-essential proteins, and form a module or sub-network (Song and Singh, 2013).

The interaction between IP3R1 and PI3K was observed under all environmental conditions. IP3R1 activation, when combined with PI3K inactivation, increases the activities of PLA2 and AA, which are decreased with a single PI3K knockdown. It was already shown that AA released by PLA2 helps to initiate apoptosis (Penzo et al., 2004). In a *Dictyostelium discoideum* chemotaxis experiment, it was also shown that cells with PI3K deficiency were more sensitive to PLA2 inhibition (Chen et al., 2007), which supports our predicted interaction between PI3K and PLA2. To this end, we hypothesized that the PI3K inactivation could be combined with the over-activation of IP3R1 to increase the activity of proteins involved in apoptosis. IP3R1 inactivation can lead to the downregulation of RGS, and reversibly, the overexpression of IP3R1 can lead to increased activity of RGS. Similar to IP3R1, RGS subtype RGS3T has been found to be involved in inducing cell death (Fisher, 2009), and it has also been found that RGS can suppress the PI3K activity downstream of the receptor (Liang et al., 2009). Therefore, the constitutive activation of IP3R1 might also negatively regulate the activity of PI3K. Systemic analysis of the most influential components and their upstream components has led us to identify novel combinations of drug targets. In various studies, combinatorial therapies have shown a decrease in drug resistance in pathogens. In combinatorial therapy, a protein associated with drug resistance can be targeted in combination with different protein of either the same or different pathway (Fischbach, 2011). Clinical trials have also suggested that the efficiency of cytotoxic drugs increases when given in combinations (Al-Lazikani et al., 2012).

In conclusion, by combining IP3R1 (activation) and PI3K (inactivation), we were able to stimulate cell death under the cell growth condition. Based on this, one can hypothesize that it might be possible that the decrease in cell proliferation with increased apoptosis as a result of this combinatorial intervention could subsequently increase the rate of clearance of tumor cells, and serve as a novel strategy for important targets associated with drug resistance. However, more laboratory validations will be required to test this hypothesis.

## 5. Acknowledgments

We would like to thank Resa Helikar for providing feedback on the manuscript.

## 6. References

Al-Lazikani, B., Banerji, U., and Workman, P. (2012). Combinatorial drug therapy for cancer in the post-genomic era. *Nature biotechnology* 30**,** 679-692.
Ashburner, M., Ball, C.A., Blake, J.A., Botstein, D., Butler, H., Cherry, J.M., Davis, A.P., Dolinski, K., Dwight, S.S., and Eppig, J.T. (2000). Gene Ontology: tool for the unification of biology. *Nature genetics* 25**,** 25-29.
Berchtold, M.W., and Villalobo, A. (2014). The many faces of calmodulin in cell proliferation, programmed cell death, autophagy, and cancer. *Biochimica et Biophysica Acta (BBA)-Molecular Cell Research* 1843**,** 398-435.
Birkeland, E., Wik, E., Mjøs, S., Hoivik, E., Trovik, J., Werner, H.M.J., Kusonmano, K., Petersen, K., Ræder, M.B., and Holst, F. (2012). KRAS gene amplification and overexpression but not mutation associates with aggressive and metastatic endometrial cancer. *British journal of cancer* 107**,** 1997-2004.

Chen, L., Iijima, M., Tang, M., Landree, M.A., Huang, Y.E., Xiong, Y., Iglesias, P.A., and Devreotes, P.N. (2007). PLA 2 and PI3K/PTEN pathways act in parallel to mediate chemotaxis. *Developmental cell* 12**,** 603-614.

Chen, W.-H., Minguez, P., Lercher, M.J., and Bork, P. (2012). OGEE: an online gene essentiality database. *Nucleic acids research* 40**,** D901-D906.

Conroy, B.D., Herek, T.A., Shew, T.D., Latner, M., Larson, J.J., Allen, L., Davis, P.H., Helikar, T., and Cutucache, C.E. (2014). Design, Assessment, and in vivo Evaluation of a Computational Model Illustrating the Role of CAV1 in CD4+ T-lymphocytes. *Frontiers in immunology* 5.

Consortium, U. (2011). Reorganizing the protein space at the Universal Protein Resource (UniProt). *Nucleic acids research***,** gkr981.

Dienstmann, R., De Dosso, S., Felip, E., and Tabernero, J. (2012). Drug development to overcome resistance to EGFR inhibitors in lung and colorectal cancer. *Molecular oncology* 6**,** 15-26.

Eisenberg, E., and Levanon, E.Y. (2013). Human housekeeping genes, revisited. *Trends in Genetics* 29**,** 569-574.

Fischbach, M.A. (2011). Combination therapies for combating antimicrobial resistance. *Current opinion in microbiology* 14**,** 519-523.

Fisher, R.A. (2009). *Molecular biology of RGS proteins.* Academic Press.

Flobak, Å., Baudot, A., Remy, E., Thommesen, L., Thieffry, D., Kuiper, M., and Lægreid, A. (2015). Discovery of Drug Synergies in Gastric Cancer Cells Predicted by Logical Modeling. *PLOS Comput Biol* 11**,** e1004426.

Flórez, A.F., Park, D., Bhak, J., Kim, B.-C., Kuchinsky, A., Morris, J.H., Espinosa, J., and Muskus, C. (2010). Protein network prediction and topological analysis in Leishmania major as a tool for drug target selection. *Bmc Bioinformatics* 11**,** 484.

Folger, O., Jerby, L., Frezza, C., Gottlieb, E., Ruppin, E., and Shlomi, T. (2011). Predicting selective drug targets in cancer through metabolic networks. *Molecular systems biology* 7**,** 501.

Ghosh, S., Matsuoka, Y., Asai, Y., Hsin, K.-Y., and Kitano, H. (2011). Software for systems biology: from tools to integrated platforms. *Nature Reviews Genetics* 12**,** 821-832.

Gottesman, M.M. (2002). Mechanisms of cancer drug resistance. *Annual review of medicine* 53**,** 615-627.

Helikar, T., Konvalina, J., Heidel, J., and Rogers, J.A. (2008). Emergent decision-making in biological signal transduction networks. *Proceedings of the National Academy of Sciences* 105**,** 1913-1918.

Helikar, T., Kowal, B., McClenathan, S., Bruckner, M., Rowley, T., Madrahimov, A., Wicks, B., Shrestha, M., Limbu, K., and Rogers, J.A. (2012). The cell collective: toward an open and collaborative approach to systems biology. *BMC systems biology* 6**,** 96.

Helikar, T., Kowal, B., and Rogers, J.A. (2013). A cell simulator platform: the cell collective. *Clinical Pharmacology & Therapeutics* 93**,** 393-395.

Helikar, T., and Rogers, J.A. (2009). ChemChains: a platform for simulation and analysis of biochemical networks aimed to laboratory scientists. *BMC systems biology* 3**,** 58.

Holohan, C., Van Schaeybroeck, S., Longley, D.B., and Johnston, P.G. (2013). Cancer drug resistance: an evolving paradigm. *Nature Reviews Cancer* 13**,** 714-726.

Huang, D.W., Sherman, B.T., and Lempicki, R.A. (2008). Systematic and integrative analysis of large gene lists using DAVID bioinformatics resources. *Nature protocols* 4**,** 44-57.

Janssens, V., and Goris, J. (2001). Protein phosphatase 2A: a highly regulated family of serine/threonine phosphatases implicated in cell growth and signalling. *Biochem. J* 353**,** 417-439.

Kanehisa, M. (2002). The KEGG database. *Silico Simulation of Biological Processes* 247**,** 103.

Kayl, A.E., and Meyers, C.A. (2006). Side-effects of chemotherapy and quality of life in ovarian and breast cancer patients. *Current Opinion in Obstetrics and Gynecology* 18**,** 24-28.

Kitano, H. (2002a). Computational systems biology. *Nature* 420**,** 206-210.

Kitano, H. (2002b). Systems biology: a brief overview. *Science* 295**,** 1662-1664.

Kobayashi, S., Boggon, T.J., Dayaram, T., Jänne, P.A., Kocher, O., Meyerson, M., Johnson, B.E., Eck, M.J., Tenen, D.G., and Halmos, B. (2005). EGFR mutation and resistance of non–small-cell lung cancer to gefitinib. *New England Journal of Medicine* 352**,** 786-792.

Kochi, N., Helikar, T., Allen, L., Rogers, J. A., Wang, Z., & Matache, M. T. (2014). Sensitivity analysis of biological Boolean networks using information fusion based on nonadditive set functions. *BMC systems biology*, *8*(1), 92.

Le Novère, N. (2015). Quantitative and logic modelling of molecular and gene networks. *Nature Reviews Genetics*.

Liang, G., Bansal, G., Xie, Z., and Druey, K.M. (2009). RGS16 inhibits breast cancer cell growth by mitigating phosphatidylinositol 3-kinase signaling. *Journal of Biological Chemistry* 284**,** 21719-21727.

Linardou, H., Dahabreh, I.J., Kanaloupiti, D., Siannis, F., Bafaloukos, D., Kosmidis, P., Papadimitriou, C.A., and Murray, S. (2008). Assessment of somatic k-RAS mutations as a mechanism associated with resistance to EGFR-targeted agents: a systematic review and meta-analysis of studies in advanced non-small-cell lung cancer and metastatic colorectal cancer. *The lancet oncology* 9**,** 962-972.

Lito, P., Rosen, N., and Solit, D.B. (2013). Tumor adaptation and resistance to RAF inhibitors. *Nature medicine* 19**,** 1401-1409.

Loscalzo, J., and Barabasi, A.L. (2011). Systems biology and the future of medicine. *Wiley Interdisciplinary Reviews: Systems Biology and Medicine* 3**,** 619-627.

Lotfi-Jam, K., Carey, M., Jefford, M., Schofield, P., Charleson, C., and Aranda, S. (2008). Nonpharmacologic strategies for managing common chemotherapy adverse effects: a systematic review. *Journal of Clinical Oncology* 26**,** 5618-5629.

Madrahimov, A., Helikar, T., Kowal, B., Lu, G., and Rogers, J. (2013). Dynamics of influenza virus and human host interactions during infection and replication cycle. *Bulletin of mathematical biology* 75**,** 988-1011.

Mason, R.P. (1999). Effects of calcium channel blockers on cellular apoptosis. *Cancer* 85**,** 2093-2102.

Mendelsohn, J. (2001). The epidermal growth factor receptor as a target for cancer therapy. *Endocrine-Related Cancer* 8**,** 3-9.

Molinelli, E.J., Korkut, A., Wang, W., Miller, M.L., Gauthier, N.P., Jing, X., Kaushik, P., He, Q., Mills, G., and Solit, D.B. (2013). Perturbation biology: inferring signaling networks in cellular systems.

Naldi, A., Carneiro, J., Chaouiya, C., and Thieffry, D. (2010). Diversity and plasticity of Th cell types predicted from regulatory network modelling. *PLoS Comput Biol* 6**,** e1000912.

Penzo, D., Petronilli, V., Angelin, A., Cusan, C., Colonna, R., Scorrano, L., Pagano, F., Prato, M., Di Lisa, F., and Bernardi, P. (2004). Arachidonic acid released by phospholipase A2 activation triggers $Ca^{2+}$-dependent apoptosis through the mitochondrial pathway. *Journal of Biological Chemistry* 279**,** 25219-25225.

Perna, D., Karreth, F.A., Rust, A.G., Perez-Mancera, P.A., Rashid, M., Iorio, F., Alifrangis, C., Arends, M.J., Bosenberg, M.W., and Bollag, G. (2015). BRAF inhibitor resistance mediated by the AKT pathway in an oncogenic BRAF mouse melanoma model. *Proceedings of the National Academy of Sciences* 112**,** E536-E545.

Pinton, P., Giorgi, C., Siviero, R., Zecchini, E., and Rizzuto, R. (2008). Calcium and apoptosis: ER-mitochondria $Ca^{2+}$ transfer in the control of apoptosis. *Oncogene* 27**,** 6407-6418.

Pruitt, K.D., Tatusova, T., and Maglott, D.R. (2007). NCBI reference sequences (RefSeq): a curated

non-redundant sequence database of genomes, transcripts and proteins. *Nucleic acids research* 35, D61-D65.

Puniya, B.L., Kulshreshtha, D., Verma, S.P., Kumar, S., and Ramachandran, S. (2013). Integrated gene co-expression network analysis in the growth phase of Mycobacterium tuberculosis reveals new potential drug targets. *Molecular BioSystems* 9, 2798-2815.

Qu, Y., Chen, Q., Lai, X., Zhu, C., Chen, C., Zhao, X., Deng, R., Xu, M., Yuan, H., and Wang, Y. (2014). SUMOylation of Grb2 enhances the ERK activity by increasing its binding with Sos1. *Mol Cancer* 13, 95.

Rachagani, S., Senapati, S., Chakraborty, S., Ponnusamy, M.P., Kumar, S., Smith, L., Jain, M., and Batra, S.K. (2011). Activated KrasG12D is associated with invasion and metastasis of pancreatic cancer cells through inhibition of E-cadherin. *British journal of cancer* 104, 1038-1048.

Rocha, C., Mendonça, T., and Eduarda Silva, M. (2013). Modelling neuromuscular blockade: a stochastic approach based on clinical data. *Mathematical and Computer Modelling of Dynamical Systems* 19, 540-556.

Rodon, J., Dienstmann, R., Serra, V., and Tabernero, J. (2013). Development of PI3K inhibitors: lessons learned from early clinical trials. *Nature reviews Clinical oncology* 10, 143-153.

Singh, P., and Singh, A. (2012). Ocular adverse effects of Anti-cancer Chemotherapy. *journal of cancer therapeutics and research* 1, 5.

Song, J., and Singh, M. (2013). From hub proteins to hub modules: the relationship between essentiality and centrality in the yeast interactome at different scales of organization.

Todd, R.G., and Helikar, T. (2012). Ergodic sets as cell phenotype of budding yeast cell cycle.

Vanneman, M., and Dranoff, G. (2012). Combining immunotherapy and targeted therapies in cancer treatment. *Nature Reviews Cancer* 12, 237-251.

Wagle, N., Van Allen, E.M., Treacy, D.J., Frederick, D.T., Cooper, Z.A., Taylor-Weiner, A., Rosenberg, M., Goetz, E.M., Sullivan, R.J., and Farlow, D.N. (2014). MAP kinase pathway alterations in BRAF-mutant melanoma patients with acquired resistance to combined RAF/MEK inhibition. *Cancer discovery* 4, 61-68.

Wang, J., Tsang, W. W., & Marsaglia, G. (2003). Evaluating Kolmogorov's distribution. *Journal of Statistical Software*, 8(18).

Wazir, U., Jiang, W.G., Sharma, A.K., and Mokbel, K. (2013). Guanine nucleotide binding protein β 1: a novel transduction protein with a possible role in human breast cancer. *Cancer Genomics-Proteomics* 10, 69-73.

Weinstein, J.N., Collisson, E.A., Mills, G.B., Shaw, K.R.M., Ozenberger, B.A., Ellrott, K., Shmulevich, I., Sander, C., Stuart, J.M., and Network, C.G.A.R. (2013). The cancer genome atlas pan-cancer analysis project. *Nature genetics* 45, 1113-1120.

West, K.A., Castillo, S.S., and Dennis, P.A. (2002). Activation of the PI3K/Akt pathway and chemotherapeutic resistance. *Drug Resistance Updates* 5, 234-248.

Wheeler, D.L., Dunn, E.F., and Harari, P.M. (2010). Understanding resistance to EGFR inhibitors—impact on future treatment strategies. *Nature reviews Clinical oncology* 7, 493-507.

Wishart, D.S., Knox, C., Guo, A.C., Shrivastava, S., Hassanali, M., Stothard, P., Chang, Z., and Woolsey, J. (2006). DrugBank: a comprehensive resource for in silico drug discovery and exploration. *Nucleic acids research* 34, D668-D672.

Yao, L., Evans, J.A., and Rzhetsky, A. (2010). Novel opportunities for computational biology and sociology in drug discovery: Corrected paper. *Trends in biotechnology* 28, 161-170.

## 7. Tables

**Table 1: Activity level ranges of environmental stimuli for cell death, growth, motility,**

**quiescence, and random environments**

| External | Death | Growth | Motility | Quiescence | Random |
|---|---|---|---|---|---|
| Extracellular matrix (ECM) | 10 - 72 | 26 - 82 | 81- 99 | 7 - 30 | 0 - 100 |
| Epidermal Growth Factor (EGF) | 3 - 15 | 72 - 97 | 29 - 83 | 43 - 56 | 0 - 100 |
| Calcium Pump (ExtPump) | 35 - 87 | 24 - 83 | 41 - 92 | 17 - 82 | 0 - 100 |
| GPCR q ligand (alpha_qL) | 13 - 58 | 18 - 78 | 17 - 74 | 4 - 84 | 0 - 100 |
| GPCR i ligand (alpha_iL) | 1 - 4 | 15 - 77 | 30 - 82 | 31 - 83 | 0 - 100 |
| GPCR s ligand (alpha_sL) | 30 - 87 | 24 - 80 | 20 - 77 | 19 - 46 | 0 - 100 |
| GPCR 12/13 ligand (alpha_1213L) | 14 - 65 | 18- 78 | 12 - 77 | 18 - 67 | 0 - 100 |
| IL1_TNF | 4 - 13 | 8 - 15 | 4 - 13 | 2 | 2 |
| Stress | 2 - 5 | 2 - 5 | 2 - 5 | 2 - 3 | 2 |

**Table 2: Housekeeping genes in the most influential components overlapped among different environmental conditions**

| Perturbation | Components | Genes | Housekeeping genes* |
|---|---|---|---|
| Inactivating | PI4K | PI4KA, PI4KB, PIK4CB | PI4KA, PI4KB |
| | PI5K | PIP5K1A, PIP5K1B, PIP5K1C | PIP5K1A |
| | ARF | ARFGAP1, ARFGAP2, ARFGAP3 | ARFGAP2, ARFGAP3 |
| | PP2A | PPP2CA | PPP2CA |
| | PI3K | PIK3CA, PIK3CB, PIK3CD, PIK3CG | PIK3C3 , PIK3CB |
| Activating | EGFR | EGFR | No |
| | IL1_TNFR | IL1B, TNFRSF1A | No |
| | TRAFS | TRAF1, TRAF2, TRAF3, TRAF4, TRAF5, TRAF6, TRAF7 | TRAF6, TRAF7 |
| | ERK | MAPK1 to MAPK15 | MAPK1, MAPK6, MAPK8, MAPK9 |
| | MEK | MAP2K1 to MAP2K7 | MAP2K1, MAP2K2, |

|   |   |   | MAP2K5 |
|---|---|---|---|
|   | PKC | PRKCA, PRKCB, PRKCD, PRKCE, PRKCG, PRKCH, PRKCI, PRKCQ, PRKCZ | No |
|   | GAB1 | GAB1 | No |
|   | SHP2 | PTPN11 | PTPN11 |

*List of housekeeping genes were obtained from (Eisenberg and Levanon, 2013).

**Table 3: Condition specific components and literature support**

| Perturbations | Environmental Condition | Associated components | Literature |
|---|---|---|---|
| Inactivating | Death | **CaM**, **RGS**, Palpha_iR | CaM and CaM-dependent signaling systems control vertebrate cell proliferation, programmed cell death and autophagy (Berchtold and Villalobo, 2014). RGS is involved in cell death (Fisher, 2009) |
| Activating | Death | **Gbg_i** (GNB), Alpha_iR | Gbg_i has been hypothesized to be involved in mTOR mediated anti-apoptotic pathways. Futhermore, it has been functionally annotated with apoptosis, cell death (Wazir et al., 2013) |
|   | Growth | PP2A | Highly regulated family of Ser/Thr phosphatase implicated in cell growth and signaling (Janssens and Goris, 2001) |
|   | Motility | KRAS, Sos | Knock-down of *KRAS* in pancreatic cancer cell lines leads to decreased motility and proliferation. The Grb2-Sos1 complex may promote cell motility, and tumerogenesis (Qu et al., 2014) |

**Table 4: Number of downstream targets of the most influential druggable components**

|   | Number of affected components | Number of affected druggable components | Number of cancer associated components | Perturbation |
|---|---|---|---|---|
| EGFR | 70 | 25 | 8 | Inactivating |
| EGFR | 24 | 13 | 3 | Activating |
| IL1_TNFR | 54 | 14 | 5 | Activating |
| Erk | 54 | 21 | 8 | Activating |
| SHP2 | 53 | 17 | | Activating |
| RKIP | 43 | 12 | 4 | Activating |
| PI3K | 42 | 17 | 7 | Inactivating |
| PP2A | 36 | 14 | 6 | Inactivating |
| PP2A | 5 | 3 | 2 | Activating |
| Ras | 30 | 13 | 5 | Activating |
| GRK | 22 | 5 | 2 | Inactivating |
| Gbg_i | 15 | 5 | 1 | Activating |
| Fak | 14 | 6 | 4 | Activating |
| Integrins | 11 | 3 | 3 | Activating |
| CaM | 8 | 5 | 2 | Inactivating |

**Table 5: Fold differences of the affected most influential component when the upstream component was perturbed.**

| Perturbed component | Affected component | Fold differences (Perturbed/WT) | | | |
|---|---|---|---|---|---|
| | | Death | Growth | Motility | Quiescence |
| IP3R1 (inactivation) | PI3K | 2.38 Fold* | 1.03 Fold | 1.04 Fold | 1.14 Fold |
| PI3K (inactivation) | IP3R1 | 5.42 Fold* | 1.18 Fold | 1.15 Fold | 1.24 Fold |

| IP3R1 (inactivation) | RGS | NSA | complete inactivation | complete inactivation | complete inactivation |
|---|---|---|---|---|---|
| RGS (inactivation) | IP3R1 | NSA | 1.21 Fold | 1.18 | 1.24 Fold |
| Gbg_i (inactivation) | CaM | NSA | NSA | complete inactivation | complete inactivation |
| CaM (inactivation) | Gbg_i | NSA | NSA | 1.30 Fold | 1.43 Fold |
| Grb2 (activation) | Ras | 7.40 Fold* | 1.32 Fold | 1.39 Fold | 2.13 Fold |
| Ras (activation) | Grb2 | 0.99 Fold | 0.97 Fold | 0.99 Fold | 1.01 Fold |
| Grb2 (activation) | Sos | 7.87 Fold* | 1.39 Fold | 1.53 Fold | 2.18 Fold |
| Sos (activation) | Grb2 | 1 Fold | 0.97 Fold | 0.99 Fold | 1.01 Fold |
| Rac (activation) | PAK | 18.41 Fold* | 5.69 Fold | NSA | NSA |
| PAK (activation) | Rac | 1.18 Fold | 1.24 Fold | NSA | NSA |

NSA = Not significantly affected (KST value < 1)
* Two fold or above change

**Table 6: Activity of affected components under single (PI3K or IP3R1) and double perturbations (PI3K and IP3R1) under the cell growth environmental condition**

| Affected components | PI3K inactivation (Single perturbation)* | IP3R1 activation (Single perturbation)* | Double perturbation* | Functional annotation** |
|---|---|---|---|---|
| Rap1 | 3.25 fold | 1.07 fold | 3.90 fold | Tumor suppressor gene |
| Ca | 1.17 fold | 1.41 fold | 1.43 fold | Calcium ion, apoptosis |
| CaM | 1.17 fold | 1.41 fold | 1.43 fold | Cell Death |
| CaMKK | 1.17 fold | 1.41 fold | 1.43 fold | Calcium ion binding, apoptosis |
| Myosin | 0.30 fold | 1.004 fold | 0.36 fold | Regulatory light chain of myosin |
| CaMK | 1.33 fold | 2.09 fold | 2.19 fold | may function in dendritic spine and synapse formation and neuronal plasticity |
| PLA2 | 0.32 fold | 1.24 fold | 0.63 fold | Tumor suppressor gene, Apoptosis |
| AA | 0.32 fold | 1.24 fold | 0.63 fold | Apoptosis |

*Compared to the activity of components in wild type.

** Functional annotations for proteins were obtained from UniProt database and literature.

**8. Figure Legends**

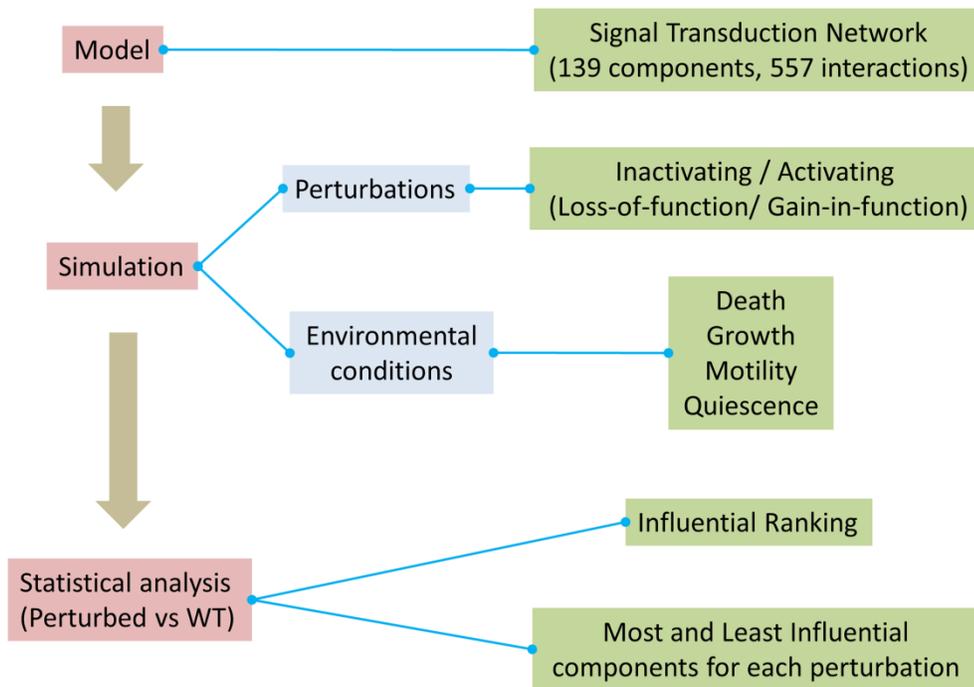

**Figure 1: Overview of the method used to assess influential components in the model.**

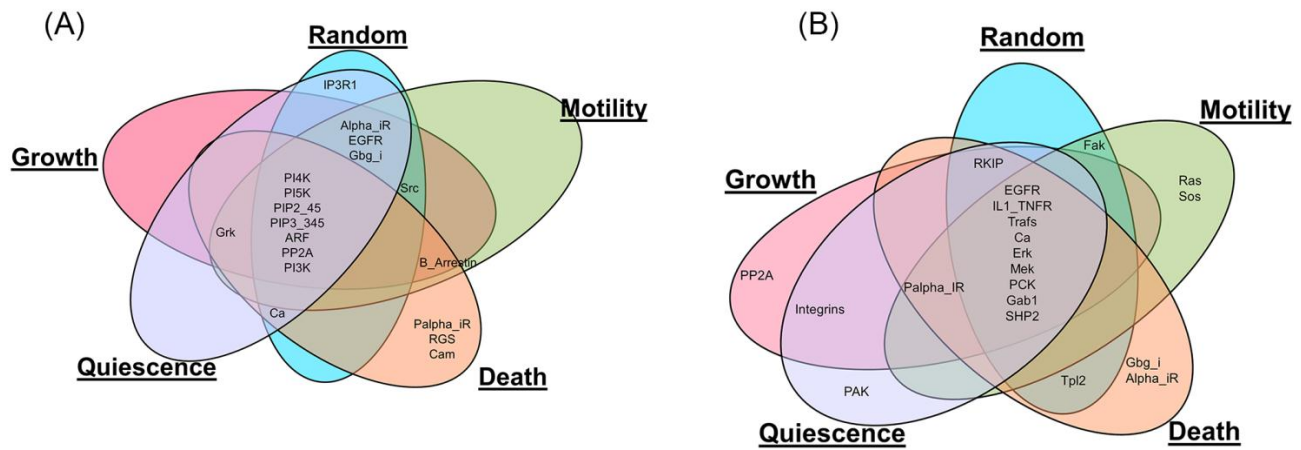

**Figure 2: Comparison of the most influential components across simulated environmental conditions.** (A) Inactivating perturbations, (B) Activating perturbations.

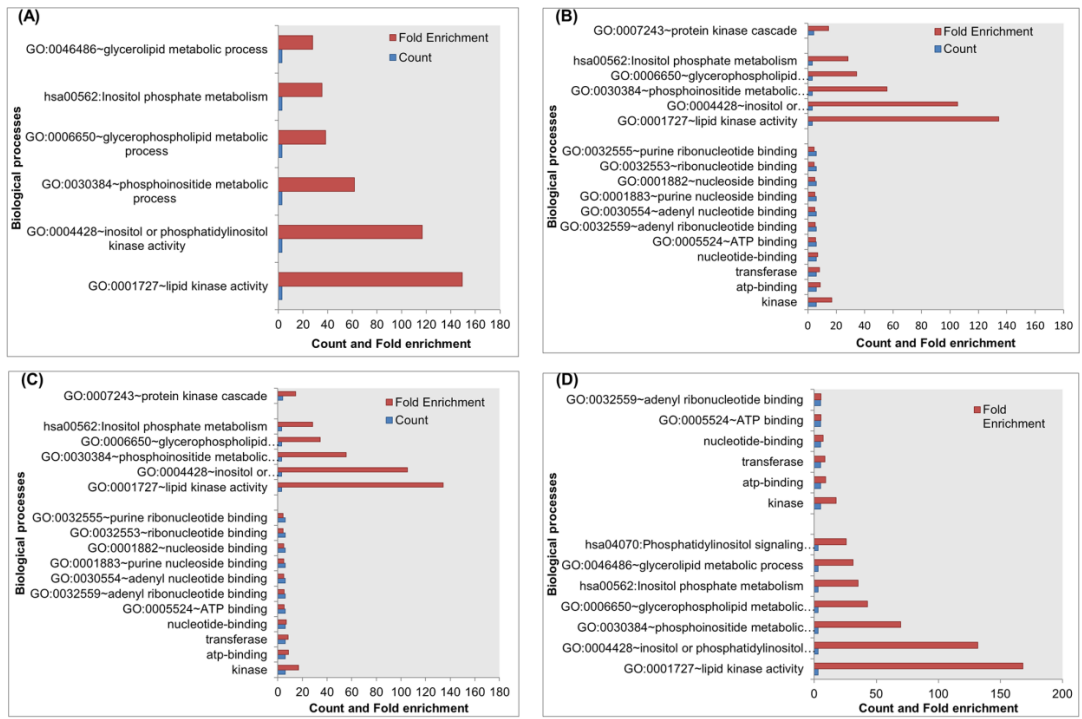

**Figure 3:** Enriched biological processes in the most influential components under environmental conditions, and inactivating perturbations.

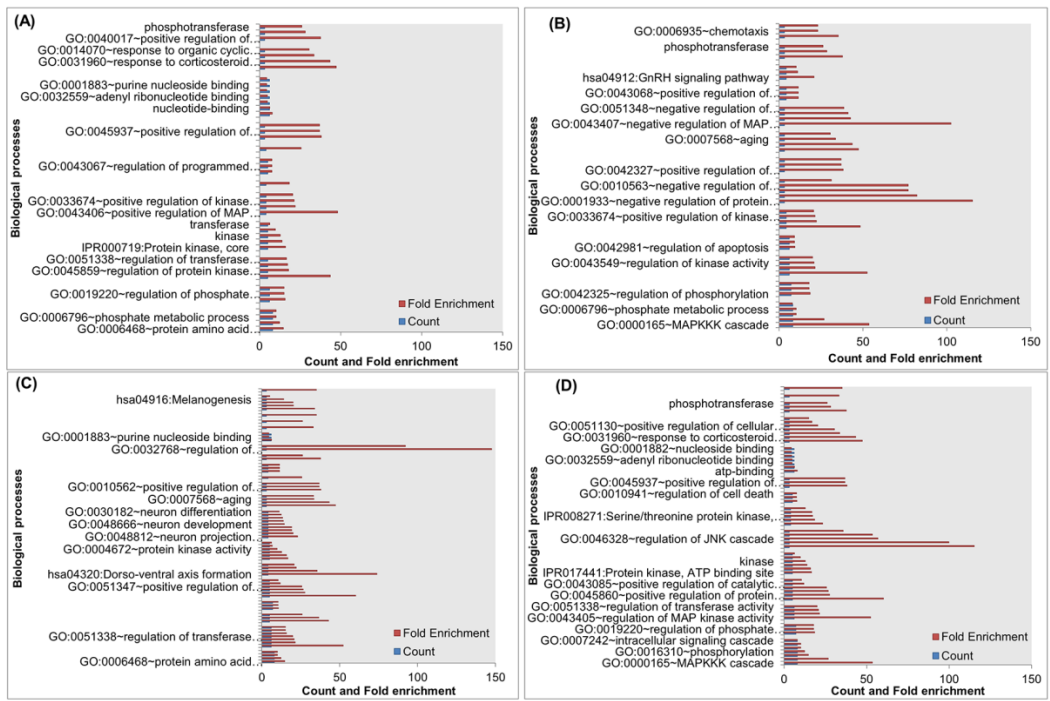

**Figure 4:** Enriched biological processes in the most influential components under environmental conditions, and activating perturbations.

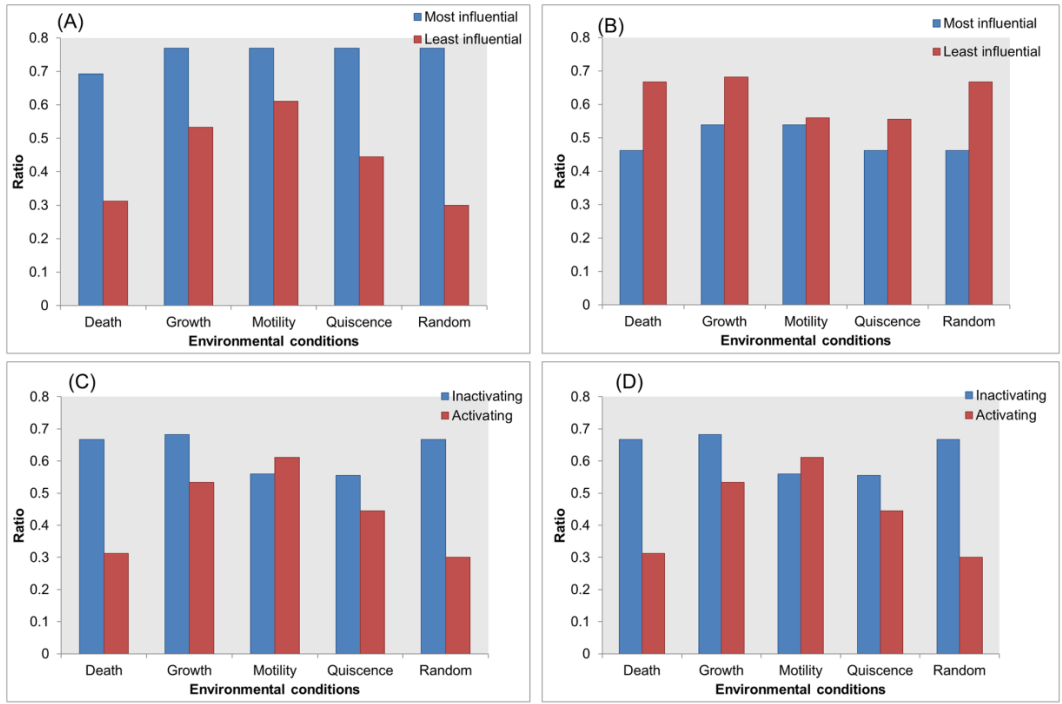

**Figure 5: Distribution of essential genes in the most influential components**. X-axis= environmental conditions, Y-axis = ratio of essential genes in total selected most or least influential components in (A) Most influential *vs* least influential components under activating perturbations (B) Most influential *vs* least influential components under inactivating perturbations (C) Essential genes in most influential under inactivating *vs* activating perturbations (D) Essential genes in least influential components under inactivating *vs* activating perturbations.

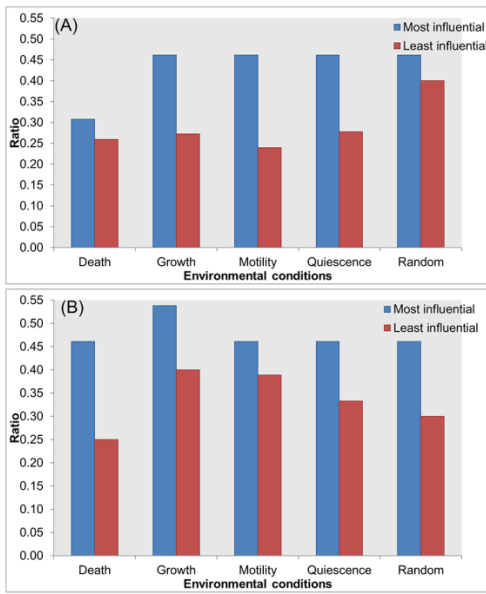

**Figure 6: Distribution of druggable proteins within the most influential vs least influential components.** (A) Inactivating perturbations, (B) Activating perturbations. X-axis = environmental conditions, Y-axis = ratio of druggable proteins in total most or least influential components.

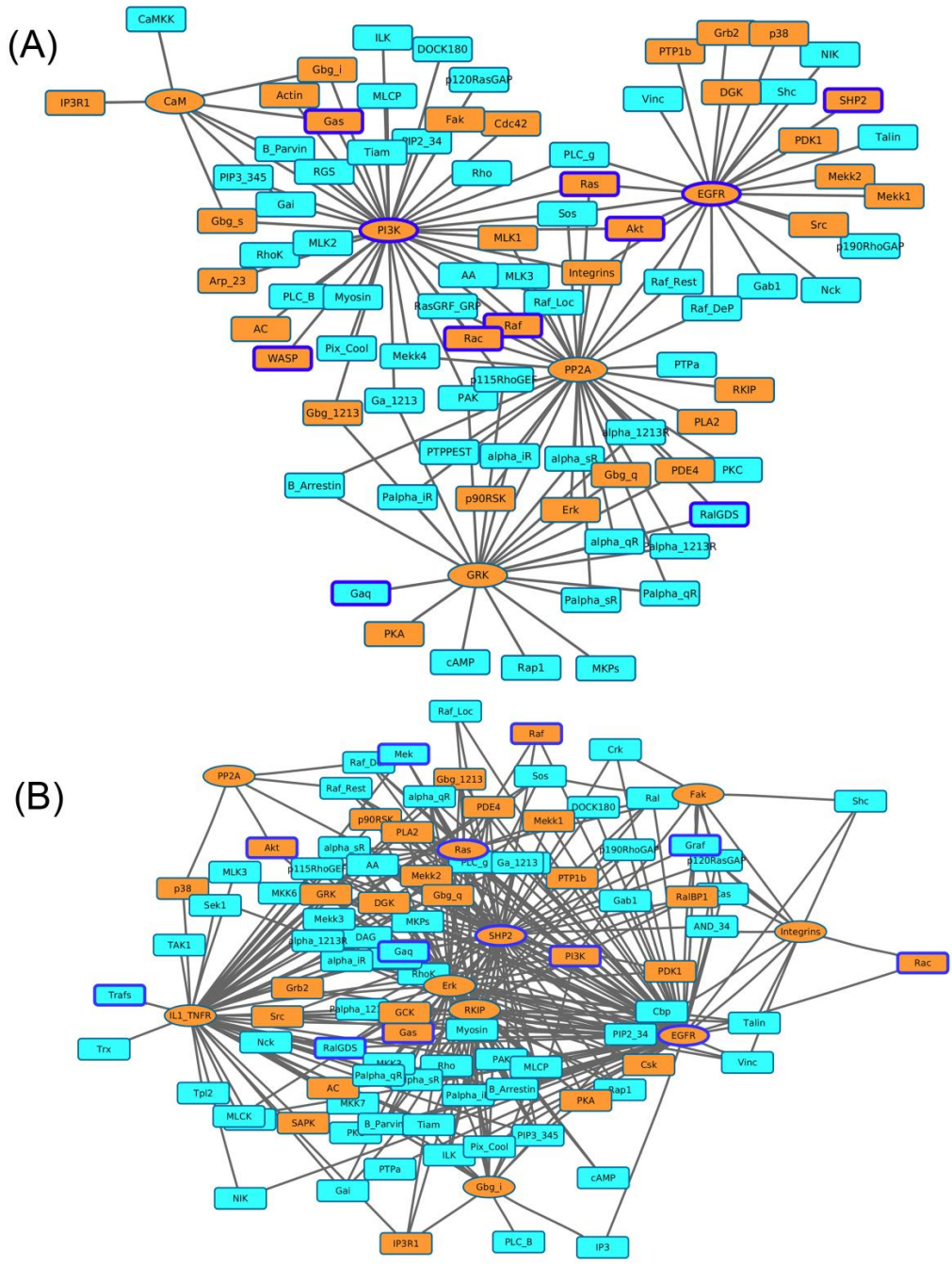

**Figure 7: Visualization of the most affected components (KST value =1) as a result of perturbing the most influential druggable components.** (A) Inactivating perturbations (B) Activating perturbations. Orange colored eclipses = most influential druggable components; squares = affected components; orange colored squares= affected druggable components; components with blue borders = experimentally found to be associated with cancer.

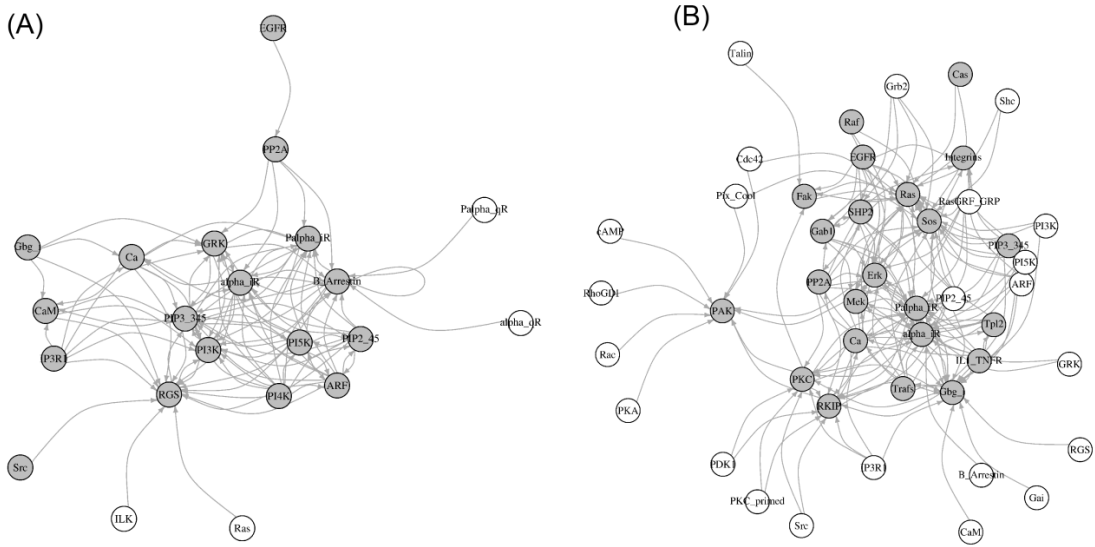

**Figure 8: Visualization of the upstream components affecting the most influential components**
(A) Inactivating perturbations (B) Activating perturbations. Grey colored nodes = the most influential components, and white colored nodes = not most influential components. The directions of arrows are from the source (upstream component) to the target (most influential components).

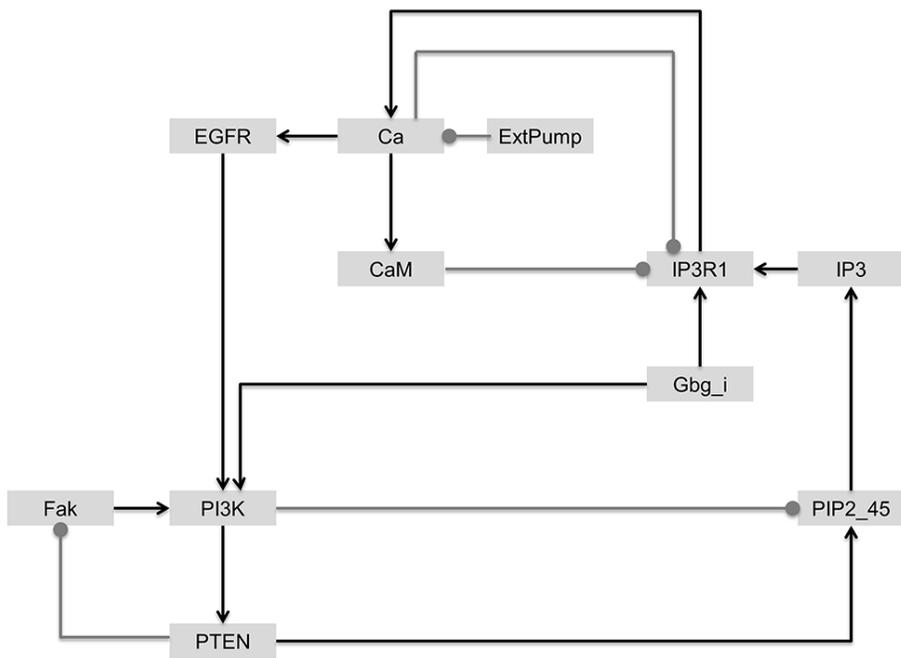

**Figure 9: The regulatory circuit connecting IP3R1 and PI3K.** Edges with arrow = activation. Edges with oval end = inhibition.